\documentclass[amsmath,amssymb,superscriptaddress,twocolumn]{revtex4-1}

\usepackage{graphicx}
\usepackage{dcolumn}
\usepackage{bm}
\usepackage{amsmath}
\usepackage{amssymb,amsthm}
\usepackage{color}
\usepackage{epstopdf} 
\usepackage[sans]{dsfont}
\usepackage{accents}
\usepackage[utf8]{inputenc}
\usepackage[T1]{fontenc}
\usepackage[english]{babel}
\usepackage{hyperref}
\usepackage{float}
\usepackage{accents}
\usepackage{amscd}
\usepackage{latexsym}
\usepackage[mathscr]{eucal}
\usepackage{enumerate}
\usepackage{hhline}
\usepackage{siunitx}

\graphicspath{{Figures/}} 

\let\originalleft\left
\let\originalright\right
\renewcommand{\left}{\mathopen{}\mathclose\bgroup\originalleft}
\renewcommand{\right}{\aftergroup\egroup\originalright}

\def\H{\ensuremath{H}}

\newcommand{\Hz}{\ensuremath{H}_0}
\newcommand{\Hrf}{\ensuremath{H}_1}
\newcommand{\He}{\ensuremath{H}_{\rm e}}
\newcommand{\Ome}[1][]{\ensuremath{\Omega_{e#1}}}
\newcommand{\om}[1][]{\ensuremath{\omega}}
\newcommand{\omo}[1][]{\ensuremath{\Omega_{0#1}}}
\newcommand{\omrf}[1][]{\ensuremath{\omega_{1#1}}}

\newcommand{\Idr}[2][]{\ensuremath{I^{#1\prime}_{#2}}}
\newcommand{\IIxyz}[5]{\ensuremath{\Idr{i #1}\Idr{j #2}#3\Idr{i #4}\Idr{j #5}}}

\newcommand{\rii}[2][2]{\ensuremath{r_{ij #2}^{#1}}}

\newcommand{\Hdip}{\ensuremath{\mathcal{H}_{\rm  dip}}}
\newcommand{\Ham}{\ensuremath{\mathcal}}
\newcommand{\Hdipo}{\ensuremath{\mathcal{H}_{\rm sec}}}
\newcommand{\HdipTOme}{\ensuremath{\mathcal{H}_{\rm ns}}}

\newcommand{\VEC}{\boldsymbol}

\newcommand{\Tns}{\ensuremath{T_1^{\rm ns}}}

\begin{document}

\title{Anomalous longitudinal relaxation of nuclear spins in CaF$_2$}

\author{Chahan M. Kropf} 
\email{chahan.kropf@physik.uni-freiburg.de}
\affiliation{Institute of Physics, University of Freiburg, Hermann-Herder-Str.~3, D-79104 Freiburg, Germany}

\author{Jonas Kohlrautz}
\affiliation{University of Leipzig, Faculty of Physics and Earth Sciences, Linn\'{e}str. 5, 04103 Leipzig, Germany}

\author{J\"urgen Haase}
\affiliation{University of Leipzig, Faculty of Physics and Earth Sciences, Linn\'{e}str. 5, 04103 Leipzig, Germany}

\author{Boris V. Fine}
\email{b.fine@skoltech.ru}
\affiliation{Skolkovo Institute of Science and Technology, 100 Novaya Str., Skolkovo, Moscow Region 143025, Russia}
\affiliation{Institute for Theoretical Physics, University of Heidelberg, Philosophenweg 12, 69120 Heidelberg, Germany}

\date{\today}

\begin{abstract}
We consider the effect of non-secular resonances for interacting nuclear spins in solids which were predicted theoretically to exist in the presence of strong static and strong radio-frequency magnetic fields. These resonances imply corrections to the standard secular approximation for the nuclear spin-spin interaction in solids, which, in turn, should lead to an anomalous longitudinal relaxation in nuclear magnetic resonance experiments. In this article, we investigate the feasibility of the experimental observation of this anomalous longitudinal relaxation in calcium fluoride (CaF$_2$) and conclude that such an observation is realistic.
\end{abstract}

\maketitle


\section{Introduction}

 In this article, we consider effective nuclear spin-spin interaction in solids in the presence of strong static and strong radio-frequency (RF) magnetic fields that can be realized in nuclear magnetic resonance (NMR) experiments. In Ref.~\cite{Kropf2012}, two of us showed that, in the above setting, the standard truncation procedure of the full magnetic dipole interaction introduced by Van Vleck~\cite{Vanvleck1948}  and Redfield~\cite{Redfield1955} must be modified in the presence of non-secular resonances associated with certain combinations of the static field, the RF field amplitude and the RF field frequency. At the non-secular resonances, significant non-secular terms appear in the effective spin-spin interaction Hamiltonian, which do not conserve the total polarization of the system along the direction of the effective magnetic field, which, in turn, leads to anomalous longitudinal relaxation. In the present article, we investigate the feasibility of observing this anomalous longitudinal relaxation in calcium fluoride (CaF$_2$).

The rest of this article is organized as follows: in Section \ref{sec:theory} we summarize the theoretical description of the non-secular resonances, then, in Section \ref{sec:Experimental constraints}, we discuss general experimental constraints, and, in Section \ref{sec:Anomalous relxation}, the anomalous longitudinal relaxation is described in the light of these constraints. Thereafter we introduce a possible experimental procedure for observing the anomalous longitudinal relaxation in Section \ref{sec:experimental protocol} and provide estimates for the relevant parameters in Section \ref{sec:Estimate}.

\section{Theory of non-secular resonances}\label{sec:theory}

 Let us consider a cubic lattice of nuclear spins 1/2 in a static magnetic field along the $z$-axis, $\VEC{H}_0 \equiv (0,0,H_0)$, irradiated by an RF field rotating in the $x$-$y$ plane with frequency $\om$, \mbox{$\VEC{H}_1(t) \equiv (H_1 \hbox{cos}(\omega t), H_1 \hbox{sin}(\omega t),0) $} and coupled by the magnetic dipole interaction. The above setting represents $^{19}$F NMR experiments in CaF$_2$.  

The full Hamiltonian of the lattice is 
\begin{equation}
\Ham{H}= \Ham{H}_{\rm z}+\Ham{H}_{\rm rf}+\Hdip
\label{Htot}
\end{equation}
where 
\begin{equation}
\Ham{H}_{\rm z}=\sum_{i<j}^N-\VEC{H}_0 \cdot \gamma\VEC{I}_i
\label{Hz}
\end{equation}
is the Zeeman term, 
 \begin{equation}
 \Ham{H}_{\hbox{\small rf}}=\sum_{i<j}^N-\VEC{H}_1(t) \cdot \gamma\VEC{I}_i
 \label{Hrf}
\end{equation}
is the RF term, and 
\begin{equation}
\Hdip =\sum_{i<j}^N\frac{\gamma^2\hbar^2}{r_{ij}^3}\; [\VEC{I}_i\VEC{I}_j-\frac{3(\VEC{I}_i \cdot\VEC{r}_{ij})(\VEC{I}_j \cdot\VEC{r}_{ij})}{r_{ij}^2}]
\label{eq:Hdip}
\end{equation}
is the full magnetic dipole interaction. Here $\VEC{I}_i \equiv (I_{ix}, I_{iy}, I_{iz})$ are the operators of the nuclear spin $1/2$ of the $i$th lattice site, $\VEC{r}_{ij} \equiv (r_{ij,x},r_{ij,y},r_{ij,z})$ are the displacement vectors between two lattice sites $i$ and $j$, and $\gamma$ is the gyromagnetic ratio. 

The characteristic time of the spin-spin relaxation induced by the Hamiltonian $\Hdip$ is to be denoted as $T_2$. In the presence of the strong static field only, it characterizes the time scale of the non-exponential relaxation of the total nuclear magnetization transverse to the direction of the static  field. The characteristic time of the relaxation due to the interaction of nuclear spins with the lattice environment not included in the Hamiltonian (\ref{Htot}) (paramagnetic impurities and phonons in CaF$_2$) is to be denoted as $T_1$. In a strong static field, $T_1$ is the decay constant of the exponential longitudinal relaxation along the direction of the static field. We further introduce the Larmor frequency $\Omega_0 = \gamma H_0$ and the nutation frequency $\omega_1 = \gamma H_1$. 

In Ref.~\citep{Kropf2012}, the regime $\Omega_0, \omega_1 \gg 1/T_2 \gg 1/T_1$ was considered. The first of these inequalities implies the separation of time scales between the fast motion due to the external fields and the slow motion due to the spin-spin interaction. The second inequality allows us to consider nuclear spins as isolated from the environment on time scales much shorter than $T_1$. The effective spin-spin interaction Hamiltonian was calculated by averaging the full interaction Hamiltonian $\Hdip$ over the fast spin motion induced by the terms $\Ham{H}_z+\Ham{H}_{\hbox{\small rf}}$. 

\begin{figure}
	\includegraphics[width=0.5\linewidth]{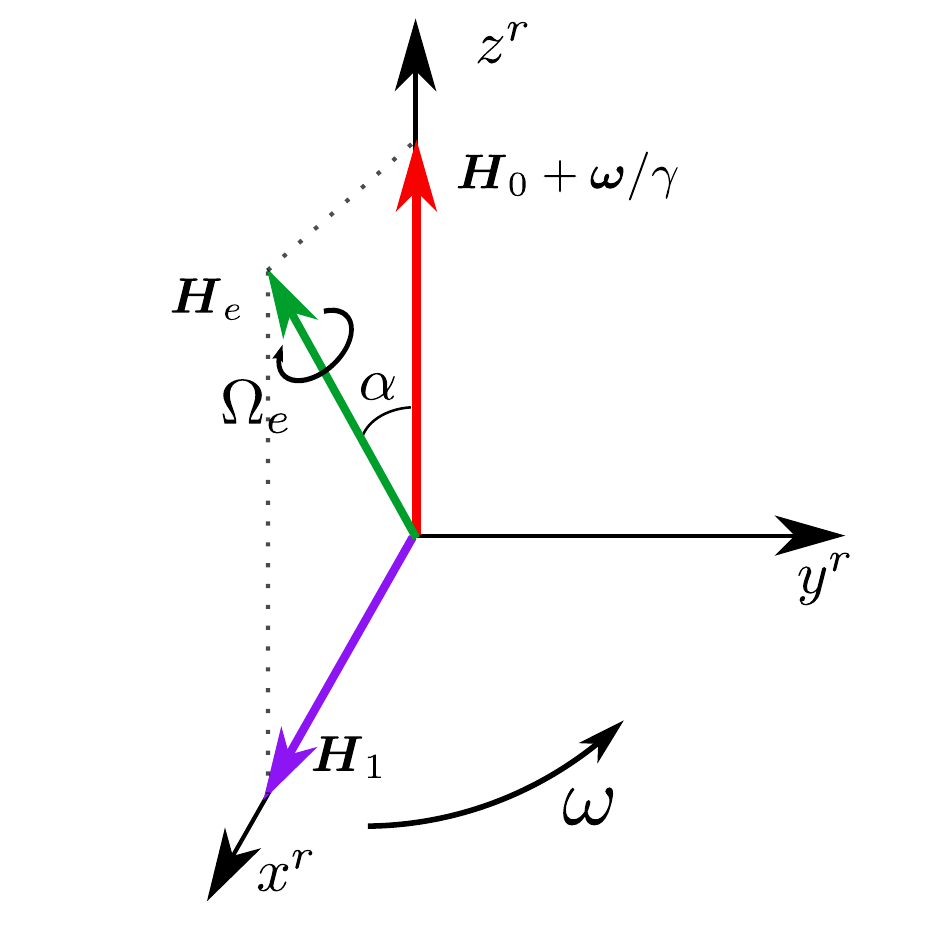}
	\caption{[Color online] Reference frame rotating at frequency $\om$ around the $z$-axis of the laboratory reference, which is parallel to the static field $\VEC{\Hz}$, and with the $x^r$-axis chosen parallel to the RF field $\VEC{\H}_1$.}
	\label{fig:frames}
\end{figure}

In the Heisenberg representation of quantum mechanics, this fast motion is described by the following equation for each of the spins~:
\begin{equation}
 \dot{\VEC{I}}_i=\VEC{I}_i\times\gamma\VEC{H}(t), 
\label{eq:Spin-eom}
\end{equation}
where $\VEC{H}(t) = \VEC{H}_0 + \VEC{H}_1(t) $. The above equation is linear in terms of the spin operators $(I_{ix}, I_{iy}, I_{iz})$, and, therefore, its predictions for quantum spins are the same as the prediction of the analogous equation for classical spin vectors with projections $(I_{ix}, I_{iy}, I_{iz})$. Despite its linearity, Eq.~(\ref{eq:Spin-eom}) would not be solvable analytically either classically or quantum mechanically for an arbitrary time-dependent $\VEC{H}(t)$. However, for the rotating RF field considered here, the solution is based on the transformation from the laboratory reference frame into the reference frame rotating around the $z$-axis with frequency $\omega$ (see Fig.~\ref{fig:frames}). (The $x^r$-axis in the rotating frame coincides with the direction of the RF field.) The equations of motion in the rotating frame have the same form as Eq.~(\ref{eq:Spin-eom}) but with $\VEC{H}(t)$ now equal to the time-independent effective magnetic field $\VEC{H}_e = (H_1, 0, H_0 + \omega/\gamma)$. As a result, the motion of each spin $\VEC{I}_i$ in the laboratory reference frame can be characterized as a Larmor precession with frequency 
\begin{equation}
    \Ome = \gamma \He=\text{sgn} (\gamma) \sqrt{\omrf^2+(\omo+\om)^2}\label{eq:Ome}
\end{equation}
around the direction of $\VEC{H}_e$ that itself rotates with frequency $\omega$ around the $z$-axis. 

The standard secular truncation due to Redfield \cite{Redfield1955} assumes that one can perform independently the averaging over the above two rotations, and hence it first cancels all terms of $\Hdip$ not commuting with $\Ham{H}_{\rm z}$, and then, it separately cancels the remaining terms not commuting with $\Ham{H}_{\rm rf}$. This leads to the effective secular interaction Hamiltonian
\begin{align}
   &\Hdipo=\sum_{i<j}^N \frac{\gamma^2\hbar^2}{r_{ij}^3} \; \frac{1}{2} \; \bigl(3\cos^2(\alpha) - 1\bigr)\; \left(1- \frac{3\rii{,z}}{\rii{}}\right)\nonumber\\ &\hspace{2cm}\times\Bigl[\Idr{iz}\Idr{jz}-\frac{1}{2}(\IIxyz{x}{x}{+}{y}{y})\Bigr]\label{eq:Hdip0}
\end{align}
where the operators $I_{ix}',I_{iy}',I_{iz}'$ are the spin projections in the reference frame rotating at frequency $\Ome$ around $\VEC{\He}$ and $\alpha$ is the angle between $\VEC{H}_e$ and $\VEC{H}_0$, which can be expressed as 
\begin{equation} \label{eq:alpha}
\alpha=\frac{\pi}{2}-\arctan\left(\frac{\omo+\om}{\omrf}\right).
\end{equation}

Independent averaging over the rotation around $\VEC{H}_e$ and the rotation of $\VEC{H}_e$  around $\VEC{H}_0$ becomes questionable when the frequencies of these two rotations are commensurate with each other.  In Ref.~\citep{Kropf2012}, two of the authors have found that, for some of these commensurate relations, averaging over the above two rotations indeed leads to additional non-secular corrections $\Ham{H}_{\rm ns}$ to  $\Hdipo$. The following five possible non-secular resonant conditions were identified: 
\begin{align}
\om&= -\Ome : & \om&= - \frac{\omrf^2+\omo^2}{2\omo} ; \label{omOme} \\ 
\om&= - 2\Ome : &  \om&= \frac{- 4 \omo + 2 \sqrt{\omo^2-3\omrf^2}}{3} ; \label{om2Ome+} \\ 
\om&= - 2\Ome : &  \om&= \frac{- 4 \omo - 2 \sqrt{\omo^2-3\omrf^2}}{3} ; \label{om2Ome-} \\ 
\om&= \frac{1}{2}\Ome : & \om&=\frac{\omo + \sqrt{4 \omo^2 + 3 \omrf^2}}{3}. \label{2ommOme+} \\
\om&=- \frac{1}{2}\Ome : & \om&=\frac{\omo - \sqrt{4 \omo^2 + 3 \omrf^2}}{3}. \label{2ommOme-}
\end{align}
When the value of $\om$ following from Eqs.~(\ref{omOme})-(\ref{2ommOme-}) is negative, this indicates that the frequency vector $\boldsymbol \omega$ is anti parallel to $\bold{\Hz}$. 

\begin{figure}
	\includegraphics[width=0.95\linewidth]{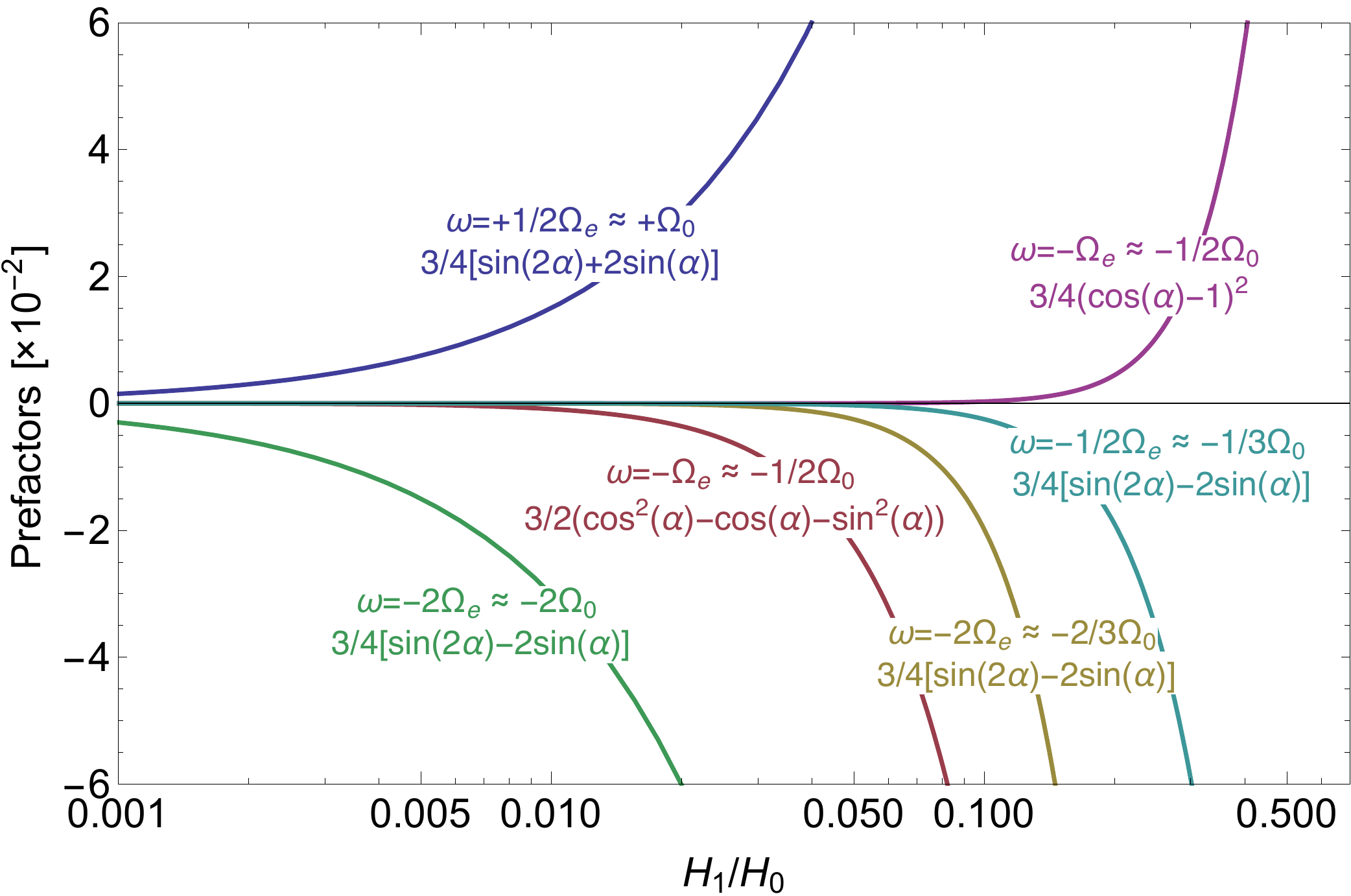}
	\caption{[Color online] Semi-logarithmic plot of the $\alpha$-dependent prefactors of the non-secular Hamiltonians $\Ham{H}_{\rm ns}$ corresponding to  the resonance conditions (\ref{omOme})-(\ref{2ommOme-}) as a function of $\Hrf/\Hz$ for $\Hrf < 0.5\Hz$. Each curve is labelled with the corresponding non-secular resonance condition and its leading-order approximation in $\Hrf/\Hz$ (indicated with $\approx$), and with the $\alpha$-dependent prefactor of the corresponding Hamiltonian $\Ham{H}_{\rm ns}$. In this article we focus on resonance condition $\om = -2\omo$ (\ref{om2Ome-}) (green). The other resonances are: $\om = -1/3\omo$ (\ref{2ommOme-}) (turquoise), $\om = +\omo$ (\ref{2ommOme+}) (blue),  $\om = -2/3\omo$ (\ref{om2Ome+}) (yellow),  and $\om=-1/2\omo$ (\ref{omOme}) (red and pink). The latter resonance conditions labels two distinct curves because the corresponding non-secular Hamiltonian $\Ham{H}_{\rm ns}$ is the sum of two parts with different prefactors, which are both shown here.} 
	\label{fig:prefactors}
\end{figure}

In the present article, for reasons to be explained shortly, we focus on the resonant condition (\ref{om2Ome-}), in which case the non-secular correction to $\HdipTOme$ is
\begin{align}
\nonumber
 &\HdipTOme = \frac{3}{4} \;[\sin (2 \alpha) - 2 \sin(\alpha)] \; \sum_{i<j}^N \frac{\gamma^2\hbar^2}{r_{ij}^3} \\
 &\;\; \times \Bigl[\frac{r_{ij,x}r_{ij,z}}{\rii{}}(\IIxyz{x}{x}{-}{y}{y})- \frac{ r_{ij,y} r_{ij,z}}{\rii{}}(\IIxyz{x}{y}{+}{y}{x})\Bigr].
\label{eq:Hdip(om=2Ome)}
\end{align}
 The form of $\Ham{H}_{\rm ns}$ for the other four resonant conditions can be found in \cite{Kropf2012}. Each of those four Hamiltonians, analogously to the one given by Eq. (\ref{eq:Hdip(om=2Ome)}), has an $\alpha$-dependent prefactor, which controls the strength of the non-secular coupling. These prefactors are plotted in Fig.~\ref{fig:prefactors}.

\section{Limit $H_1 \ll \Hz$}\label{sec:Experimental constraints}
 
 From the theoretical perspective, the non-secular resonance conditions (\ref{omOme})-(\ref{2ommOme-}) are best tested when both $H_0$ and $H_1$ are of the same order of magnitude. However, these magnetic fields are subject to the following constraints in a typical NMR setting: On the one hand, the magnetization induced voltage is proportional to $H_0^2$, which means that $H_0$ should not be too small for a given noise level. We assume that $H_0 \geq \num{2500}$~G is needed for the usual signal detection. On the other hand, it is difficult to generate a continuous RF field with an amplitude greater than $100$~G due to the associated heating. We therefore assume that $H_1 \leq 100$~G. The above considerations leave us with the condition $H_1/H_0 \ll 1$.
In such a case, $\HdipTOme \ll \Hdipo$. Nevertheless, the effect of $\HdipTOme$ may still be observable because $\HdipTOme$ does \emph{not} conserve the projection of the total nuclear polarization on the direction of $\VEC{H}_e$, while $\Hdipo$ does. As a result, $\HdipTOme$ leads to an anomalous longitudinal relaxation of the total nuclear magnetization. Whether such an anomalous relaxation is observable experimentally is the focus of the remaining part of this article.

It was obtained in Ref.~\cite{Kropf2012}, and illustrated in Fig.~\ref{fig:prefactors}, that the non-secular Hamiltonians corresponding to the different non-secular resonances~(\ref{omOme})-(\ref{2ommOme-}) are of different order in $\Hrf/\Hz$ for $\Hrf/\Hz \ll 1$. This order is linear for the resonances (\ref{om2Ome-}) and (\ref{2ommOme+}), and higher than linear for the rest, which means that the non-secular Hamiltonians for resonances (\ref{om2Ome-}) and (\ref{2ommOme+}) are the largest and hence the most promising for the experimental detection. The second of the above resonances, however, has frequency $\om \approx \omo$, while, in our sign convention, the fundamental Larmor frequency of NMR experiments is $-\omo$. Therefore, an excitation coil that produces a genuine circularly polarized RF field with $\om = + \omo$ would be needed (note that any component at $-\omo$ could easily saturate the spin system).


The above considerations leave us with the resonant condition (\ref{om2Ome-}) as the most promising one for detecting the anomalous longitudinal relaxation.

The condition $H_1/H_0 \ll 1$ implies two more simplifications for an experiment:  (1) Away from $\omega = \pm\Omega_0$, it allows us to apply the results obtained with a rotating RF field  summarized in Section~\ref{sec:theory} to the case of a linearly polarized RF-field, because the effect of the counter-rotating component averages out. (2) In the leading order, the nonsecular resonant conditions are independent of $H_1$,  which implies that these conditions are insensitive to the inhomogeneity of $H_1$ caused by a realistic RF coil. The resulting anomalous relaxation rate may still vary across the sample, but this would not cancel the overall effect. 

Finally, we would like to mention here that the inhomogeneity of $\Hz$ over the sample, $\Delta \Hz$,  is another factor controlling the accuracy of the nonsecular resonances. However, even when assuming a large inhomogeneity of $\Delta \Hz/\Hz = 10^{-5}$, we obtain for $\Hz = 10^4$~G a value of $\Delta \Hz = 0.1$~G, which is still quite tolerable, because it is much smaller than that of a typical local field, $h_{\rm loc} \sim 1$~G, with which nuclear spins act on each other (see section \ref{sec:Estimate}). As explained in Ref.~\cite{Kropf2012}, the nonsecular resonant condition is defined with accuracy $\Delta \om = \pm \gamma h_{\rm loc}$.

 \section{Anomalous longitudinal relaxation}\label{sec:Anomalous relxation}
 
 The dipolar spin-spin interaction (\ref{eq:Hdip}) can be understood as each spin $\VEC{I}_i$ interacting with a local field $\VEC{h}_{i, {\rm loc}}$ created by all the other spins so that $\Hdip = \sum_{i} \gamma \hbar \VEC{I}_i \cdot \VEC{h}_{i, {\rm loc}}$. Analogously, the secular (\ref{eq:Hdip0}) and the non-secular (\ref{eq:Hdip(om=2Ome)}) effective Hamiltonians can be expressed in terms of local fields $\VEC{h}_{i, {\rm loc}}^{\rm sec}$ and $\VEC{h}_{i, {\rm loc}}^{\rm ns}$, respectively. The condition $H_1/H_0 \ll 1$ implies, as discussed in the previous section, that $h_{i, {\rm loc}}^{\rm ns} \ll h_{i, {\rm loc}}^{\rm sec}$. For the estimates to follow below we need only to consider the average value of the local fields rather than the site to site fluctuations and therefore we drop the subscript $i$ in $h_{i, {\rm loc}}$.
 
  The transverse decay time associated with the secular term can then be estimated as the inverse root-mean-squared (rms) value of the local field, $T_2 = 2\pi \gamma^{-1}\left<(\VEC{h}_{\rm loc}^{\rm sec})^2\right>^{-1/2}$. 
 
 The anomalous longitudinal relaxation time cannot be similarly estimated from the inverse rms value of $\VEC{h}_{{\rm loc}}^{ns}$, because, given $h_{{\rm loc}}^{\rm ns} \ll h_{{\rm loc}}^{\rm sec}$ , the effect of $\VEC{h}_{{\rm loc}}^{\rm ns}$ is averaged out in the leading order by the faster motion induced by  $\VEC{h}_{ {\rm loc}}^{\rm sec}$. In the next order of perturbation theory, one expects the anomalous longitudinal relaxation to become exponential with a time constant that can be estimated as  
\begin{equation}\label{eq:T1ns}
	\Tns = \frac{<({\VEC{h}_{{\rm loc}}^{\rm sec}})^2>^{1/2}}{\gamma <(\VEC{h}_{{\rm loc}}^{\rm ns})^2>} \equiv \frac{h_{\rm loc}^{\rm sec}}{\gamma (h_{\rm loc}^{\rm ns})^2}.
\end{equation}
This situation is similar to the case of exchange narrowing \cite{Anderson1953}.

The above anomalous relaxation causes the magnetization parallel to $\VEC{H}_e$ to decay to zero. Simultaneously, the usual spin-lattice relaxation drives the magnetization along $\VEC{H}_0$ towards the equilibrium value $M_0$. Given $H_1/H_0 \ll 1$, $\VEC{H}_e$ is nearly parallel to $\VEC{H}_0$. This means that the equation for the combined spin-lattice and anomalous longitudinal relaxation for the $z$-projection of the total magnetization (parallel to $\VEC{H}_0$) can be written \cite{Slichter1990,Goldman1970} as:
\begin{align}
	\dot{M_z} = -\frac{M_z}{T_1^{\rm ns}} - \frac{M_z-M_0}{T_1}.
\end{align}
The new equilibrium value is thus 
\begin{align}
M_z = M_0 \, \frac{T_1^{\rm ns}}{T_1^{\rm ns}+T_1}.
\end{align}
Therefore, when $T_1^{\rm ns} \leq T_1$, 
the equilibrium magnetization at the nonsecular resonance becomes significantly smaller than away fron the resonance.
In particular, in the limit $T_1^{\rm ns} \ll T_1$, one has $M_z\approx M_0 \,T_1^{\rm ns}/T_1 \ll M_0$. 

It follows that the key parameter for the observation of the anomalous longitudinal relaxation is $T_1^{\rm ns}/T_1$. The smaller this ratio, the more pronounced the consequence of the anomalous relaxation. The value of the anomalous relaxation time $T_1^{\rm ns}$ for CaF$_2$ will be estimated in Section~\ref{sec:Estimate}. Note that in the regime $\Hrf\ll\Hz$,  $T_1^{\rm ns}\propto \Hz^2/\Hrf^2$ and thus the larger the RF field amplitude, the shorter $T_1^{\rm ns}$.
 
\section{Experimental procedure}\label{sec:experimental protocol}

We here propose a (double-resonance) experimental procedure in five steps for observing the anomalous longitudinal relaxation~:
\begin{enumerate}
\item Polarizing the system in the static magnetic field $\Hz$
\item Turning on the linearly polarized RF field $\VEC{H}_1$ at frequencies $\omega$ close to $2\Omega_0$.
\item Waiting for a time $\tau \geq T_1^{\rm ns}$ .
\item Turning off the RF field
\item Applying a $\pi/2$-pulse (at $\om = \omo$) and recording the initial intensity of the free-induction decay signal  
\end{enumerate} 
 As follows from previous section, we expect that, for waiting times $\tau \geq T_1^{\rm ns}$, the measured total longitudinal polarization (initial value of the free-induction decay signal in step 5) for $\omega = 2\Omega_0$ should be significantly reduced as compared to the off-resonance situation. The width of the nonsecular resonance condition is expected to be of the order of $\gamma h_{{\rm loc}}^{\rm sec}$.

\section{Estimates of the anomalous longitudinal relaxation time for CaF$_2$}
 \label{sec:Estimate}
 
 Here, we estimate $T_1^{\rm ns}$ for two sets of the static and the RF fields: one we believe to be rather optimistic, $\Hz^{(1)}=\num{2500}$~G and $\Hrf^{(1)} = 100$~G, and one more conservative, $\Hz^{(2)}=\num{10 000}$~G and $\Hrf^{(2)} = 30$~G. We will further assume a longitudinal relxation time of $T_1 = 80$~s, which approximately corresponds to the value measured at $20$~K in Ref.~\cite{Meier2012}. 

\begin{table}[b!]
\begin{tabular}{c @{\hskip 3ex} l l l  @{\hskip 3ex} l l l}
\hline \hline
~ & \multicolumn{3}{l}{Optimistic choice} & \multicolumn{3}{l}{Conservative choice} \\ 
\hline
$\Hz$~[G] & \multicolumn{3}{c}{\num{2500} ($\approx$\num{10.0}~MHz)} & \multicolumn{3}{c}{\num{10000} ($\approx$\num{40.1}~MHz)} \\ 

$\Hrf$~[G]& \multicolumn{3}{c}{100 ($\approx$\num{0.4}~MHz)} & \multicolumn{3}{c}{30 ($\approx$\num{0.1}~MHz)} \\ 

$\pi-\alpha \approx \frac{\Hrf}{\Hz}$ & \multicolumn{3}{c}{0.04} & \multicolumn{3}{c}{0.001} \\ 

$T_1$~[s] & \multicolumn{3}{c}{80} & \multicolumn{3}{c}{80} \\ 
\hline 
$\Hz$ orientation & [100] & [110] & [111] & [100]  & [110] & [111] \\ 
\hline 
$h_{loc}^{\rm sec}$~[G] & $3.0$ & $1.8$ & $1.2$ & $3.0$ & $1.8$ & $1.2$ \\ 

$h_{loc}^{\rm ns} $~[G] & $0.07$ & $0.13$ & $0.14$ & $0.005$ & $0.010$ & $0.011$ \\ 

$T_{1}^{\rm ns}$~[s] & $0.024$ & $0.0042$ & $0.0024$ & $4.8$ & $0.72$ & $0.39$ \\ 
\hline \hline
\end{tabular} 
\caption{Key experimental parameters for observing the anomalous longitudinal relaxation in CaF$_2$ for an assumed optimistic and conservative choice of external magnetic fields. $T_1^{\rm ns}$ was obtained using Eq.~(\ref{eq:T1ns}). Working with Gaussian units, we used for the gyromagnetic ratio of $^{19}$F $\gamma~=~25166.2$~rad~s$^{-1}$~Oe$^{-1}$ and for the lattice constant $a = 2.72 \times 10^{-8}$~cm.}
\label{tab:experimental param}
\end{table}

The estimates of $\Tns$ using Eq.(\ref{eq:T1ns}) are summarized in Table \ref{tab:experimental param}. Since the secular and non-secular Hamiltonians depend on the orientation of $\VEC{H}_0$ with respect to the crystal lattice of CaF$_2$, we considered three standard orientations of $\VEC{H}_0$:~$[100]$, $[110]$, and $[111]$. As can be seen in Table~\ref{tab:experimental param}, the $[111]$ orientation is the most favourable because, for this orientation, the nearest-neighbors couplings in the secular term (\ref{eq:Hdip0}) (for which one then has $1-3r^2_{ij,z}/r^2_{ij} = 0$) are suppressed. The most favourable combination among those listed in Table~\ref{tab:experimental param} is: $\Hz^{(1)}=2500$~G and $\Hrf^{(1)} = 100$~G, with the static field along the $[111]$ direction. The anomalous relaxation time $T_1^{\rm ns}\approx 0.0024$~s is then about four orders of magnitude shorter than $T_1 = 80$~s. Under these conditions, an irradiation time of about $\tau \approx 0.0048$~s may already be sufficient to detect the effect of the non-secular resonances. The least favourable case is given by the more conservative variant: $\Hz^{(1)}=\num{10000}$~G, $\Hrf^{(1)} = 30$~G, and the $[100]$ orientation. The anomalous relaxation time $T_1^{\rm ns}\approx 4.8$~s is then still about one order of magnitude shorter than $T_1$. Thereby, the calculated $T_1^{ns}$ are all smaller than $T_1$ and we conclude that the anomalous longitudinal relaxation is detectable. 

\section{Conclusion}\label{sec:conclusion}

We proposed a nuclear magnetic resonance experiment to verify the existence of the anomalous longitudinal NMR relaxation caused by non-secular resonances predicted in \citep{Kropf2012}. Given the estimates presented in Table \ref{tab:experimental param}, we conclude that the experimental observation of the anomalous longitudinal relaxation is feasible in CaF$_2$.

\bibliographystyle{apsrev4-1}
\bibliography{bibliography}

\end{document}